\documentclass[extra,mreferee]{gji} 
\usepackage{epsfig} 
\usepackage{lscape}
\usepackage{setspace}
\usepackage{amsmath,amsfonts}
\newcommand{\field}[1]{\mathbb{#1}} 
\newcommand{\myarg}[1]{(\mathbf{#1})}
\newcommand{\bhat}[1]{\mathbf{\hat{#1}}}
\title[Energy partition in the seismic coda]{Energy partition 
 of seismic coda waves in layered media: theory and application to  Pinyon Flats Observatory}
\author[L. Margerin et al.]{L. Margerin$^1$, M. Campillo$^2$,   B.A. Van Tiggelen$^3$,  R. Hennino\thanks{formerly at L.G.I.T}, \\
   $^1$ Centre Europ\'een de Recherche et d'Enseignement  de G\'eosciences de l'Environnement, \\ Universit\'e Aix Marseille, 
C.N.R.S., \\ Europ\^ole M\'editerran\'een de l'Arbois, BP 80, 13545  Aix en Provence,  France
    \\
 $^2$ Laboratoire de G\'eophysique Interne et Tectonophysique, \\ Universit\'e
 Joseph Fourier, C.N.R.S., \\ B.P. 53,
38041 Grenoble, France  \\
 $^3$  Laboratoire de Physique et Mod\'elisation de la Mati\`ere Condens\'ee, \\
Universit\'e Joseph Fourier, C.N.R.S.,  \\
Maison des Magist\`eres, 
 25 avenue des Martyrs, \\
 B.P. 166,  38042 Grenoble, France}

\date{\today}

\begin{document}

\maketitle

\begin{summary}
We have studied the partition of shear, compressional  and kinetic energies in the coda
of ten earthquakes recorded on a dense  array, located
at Pinyon Flats Observatory (PFO), California.  Deformation energies are estimated by measuring 
finite differences of the wavefield components. 
We have thoroughly studied  the validity and stability of this technique for the PFO data
 and obtained reliable measurements in the 5-7 Hz frequency band. 
  We observe a clear stabilization of the shear to compressional ($W^s/W^p$)
energy ratio in the coda, with an average value of about 2.8, which is smaller by a factor 2.5
than the  expected ratio  at the surface of a homogeneous  Poisson half-space. The ratio between the vertical and horizontal
kinetic energies ($V^2/H^2$) can be measured from 5 to 25Hz and shows an abrupt transition
from 0.1 in the 5-10Hz  band, to about 0.8 in the 15-25Hz band. These measured values are in sharp
contrast with the theoretical prediction, around 0.56, for equipartitioned elastic waves in a homogeneous half-space.  
To explain these observations, we have developed a theory of equipartition in a layered elastic half-space. 
Using a rigorous spectral decomposition of the elastic wave equation,
we define equipartition as a white noise distributed over the complete set of
eigenfunctions. This definition is shown to agree with the 
standard physical concepts in canonical cases. The theory predicts that close to the resonance frequency of a low-velocity
layer, the ratio between shear and compressional energies strongly decreases.
Using a detailed  model of the subsurface at PFO, this counterintuitive result is
found to be in good qualitative and quantitative agreement with the observations.
Near the  resonance frequency of the low-velocity structure, the drop  of the energy ratios $W^s/W^p$ and $V^2/H^2$   
is controlled by the change of ellipticity of the Rayleigh wave and the large contribution of the fundamental Love mode. At higher frequencies,
the interplay between Rayleigh and Love modes  trapped in shallow low-velocity layers is responsible for the abrupt
increase of the kinetic energy ratio. Our study demonstrates that the partition of  energy in the seismic coda contains 
information on the local geological structure.

\end{summary}
\section{Introduction}
Almost forty years ago, \citet[]{aki69} gave the first explanation of the long tail of
seismograms.  The basic idea is that the diffraction of seismic waves by the heterogeneous
crust is at the origin of the late arrivals also known as coda waves. Six years later, \citet{aki75} proposed 
the first quantitative explanation for the algebro-exponential decay of the coda with time. Two models were proposed: (1) The single scattering
approximation in which seismic waves are deflected at a single scattering site during their propagation
between source and receiver; (2) The diffusion approximation which basically describes the random walk 
 of the seismic energy in the scattering medium. Although conceptually important, these two models
were not sufficiently accurate to model successfully the space time distribution of the energy in the coda.
This difficulty was partly  overcome when \citet{wu85} introduced the acoustic Radiative Transfer
Equation (RTE). Mode conversions between {\it P} and {\it S} waves were later introduced on a phenomenological basis
 by \citet{zeng93} and \citet{sato94}.
The RTE takes into account arbitrary orders of scattering in a rigorous manner,
and  describes the behavior of intensity in a complex medium beyond
the models of single scattering and diffusion. Recently the accuracy of the RTE has been demonstrated
by comparisons with full-wave simulations in both acoustic \citep{wegler06} and elastic \citep{przybilla06} media. 
While the introduction
of the RTE in seismology was an important progress, it  did not  resolve the debate initiated by
 \citet{aki75}  on the origin of the coda:  is it dominated by single or multiple scattering? This question is of both fundamental
 and practical interest: it is  only when the propagation regime is known that it  becomes possible to separate scattering from absorption. 


In the nineties, the RTE for polarized elastic waves in an
inhomogeneous medium has been derived independently by \citet{weaver90} 
 and \citet{ryzhik96}. A particularly interesting outcome of the RTE is the predicted
stabilization of the shear ($W^s$) to compressional ($W^p$)  energy ratio ($W^s/W^p$) at large lapse time in the coda, independent
of the properties of the scatterers. This property is known as equipartition \citep{weaver90,ryzhik96}
 and was originally proposed by \citet{weaver82} to describe diffuse waves, based on statistical
physics arguments. Equipartition implies  that while the compressional and shear energy  decrease exponentially with time, their ratio
$W^s/W^p$ tends to a constant. This result was numerically confirmed by 
\citet{margerin00} who found that this behavior can occur after a few scattering events only.
 A  stabilization of the  energy ratio $W^s/W^p$ was  observed in Mexico by \citet{shapiro00} and
was found to be in excellent quantitative agreement with an equipartition theory for Rayleigh, Love and body
waves by \citet{hennino01}. The very rapid stabilization of energy ratios observed in Mexico is a clear proof that the
propagation of short period elastic waves in tectonically active regions is dominated 
 by multiple scattering. 

In this paper, we report the stabilization of the ratio $W^s/W^p$,  
and the ratio between  vertical and horizontal  kinetic  energies ($V^2/H^2$),
 in the coda of short period seismic waves. The data were collected during 
a temporary experiment conducted at the  Pinyon Flats Observatory (PFO) in California \citep{fletcher90,vernon98},
where a small aperture array was installed during six months in 1990. The observed stabilization
ratio is around 2.8 which is about 2.5 times smaller than the value expected at the surface of an
elastic Poisson half-space \citep{hennino01}. To explain this observation we have
developed an equipartition theory for elastic waves in a stratified medium. Our approach
is based on a spectral decomposition of the elastic operator developed in section  2, which involves generalized
eigenfunctions. This technique circumvents the  artificial ``locked-mode'' approximation used by \citet{hennino01}
and is in much closer agreement with the situation in the Earth.
 Equipartition can then be defined mathematically as a white noise in the
modal space. Using this formal definition, the theoretical results of \citet{weaver90} and  \citet{hennino01}
are recovered and  generalized to stratified media (section 3). The method avoids 
mode counting arguments and treats on the same footing surface and body waves.  
The data analysis and the measurements are presented in section 4. Particular
attention has been paid to the stable measurements of partial derivatives
of the wavefield. In section 5, we show that  equipartition theory in a stratified
model of the crust at PFO as proposed by \citet{fletcher90}  is in  good agreement with the observations.  
In section 6 we conclude and give directions for future research.

\section{Mathematical preliminaries}
In this part we develop formally the spectral decomposition of the elastic wave operator
in a stratified half-space. We explain how a  complete  family of generalized eigenvectors of the
elastic wave equation can be constructed, based on scattering theory. The term ``generalized'' emphasizes the fact that the
eigenfunctions are not square integrable, like  plane waves. 
A key result is that any square integrable vector field can be  expanded over the set of eigenvectors of the
elastic wave equation, which is
an important  generalization of the Fourier theorem. Detailed calculations and validation of the
theory are presented in \citet{margerin08}. 

\subsection{General properties of the elastic wave operator}
In this part, we summarize  results of \cite{dermenjian85} and \cite{secher98}. 
We consider the following elastic wave equation in $ \field{R}^3_-$, which denotes the semi-infinite medium  $z<0$.
The  Lam\'e parameters $\lambda$, $\mu$, and the density $\rho$ are sufficiently well-behaved functions of the position
vector $\mathbf{x}$. For instance, we allow these functions to be discontinuous across smooth boundaries, but we
assume that they are bounded away from zero.
The elastodynamic equation reads:
\begin{equation}
\partial_{tt} \mathbf{u}(t) + \mathbf {A u}(t) = 0,
\end{equation}
where the operator of elasticity acts on the vector wavefunctions as follows:
\begin{equation}
 \mathbf{A u}_i\myarg{x} = -\frac{1}{\rho\myarg{x}} \partial_j \left( \lambda\myarg{x} \partial_k u_k\myarg{x} \delta_{ji} + 
                                     2 \mu\myarg{x} \epsilon_{ij}(\mathbf{u})\myarg{x} \right), 
\end{equation}
where $\epsilon_{ij}$ is the usual strain tensor:
\begin{equation}
 \epsilon_{ij}(\mathbf{u})\myarg{x} = \frac{1}{2}\left(\partial_i u_j\myarg{x} + \partial_j u_i\myarg{x} \right).
\end{equation}
According to Dermenjian and Guillot (1988), the operator $\mathbf{A}$ is  positive and self-adjoint in the
Hilbert space with scalar product:
\begin{equation}  \label{scprod} 
 \langle \mathbf{u} , \mathbf{v} \rangle = \int\limits_{\field{R}^3_-}\!\! \rho\myarg{x} \mathbf{u}^{\star} \cdot \mathbf{v} d^3x .
\end{equation}
We will not dwell on the precise definition of the domain of $\mathbf{A}$. Physically speaking, the choice of the domain
ensures that the wavefunctions have finite energy. The spectrum of $\mathbf{A}$, i.e. the set of generalized
 eigenvalues of the operator, is the positive real axis and we assume
that the Lam\'e parameters  and the density  are sufficiently well-behaved so that the spectrum is absolutely continuous \citep{reed80}.
In what follows, we briefly explain how to build a complete set of eigenvectors in stratified media using general 
results from scattering theory in 3-D \citep{reed79,ramm86}.  
To carry out this task, we first need a basis of eigenvectors for the elastic operator
 in infinite homogeneous space.   
\subsection{Eigenvectors of the elastic operator in homogeneous space}
In a  homogeneous infinite space the elastic operator  takes the well-known form:
\begin{equation}
 \mathbf{A_0u} = -\frac{\lambda_0 + 2 \mu_0}{\rho_0} \nabla(\nabla \cdot \mathbf{u}) + 
              \frac{\mu_0}{\rho_0} \nabla\times\nabla\times \mathbf{u}.   \label{A0}
\end{equation}
One can verify that  plane $P$ waves:
\begin{equation} \label{ep}
 \mathbf{e}^p_{\mathbf{p}}(\mathbf{r})  = \frac{\mathbf{\hat{p}}^p}{\rho_0^{1/2} (2 \pi)^{3/2}} e^{i\mathbf{p}\cdot\mathbf{r}} 
\end{equation}
are eigenvectors of $\mathbf{A_0}$ with generalized eigenvalue $p^2 \alpha_0^2$, where $\alpha_0$ denotes the $P$ wavespeed, 
$\mathbf{\hat{p}}^p$ is a unit vector in the direction of $\mathbf{p} =(p_x,p_y,p_z)$. The term generalized is used to emphasize
the fact that the eigenfunctions are not square integrable. The notational convention used in  Equation (\ref{ep})
is as follows: the superscript refers to the wave type and the subscript serves as a label for the eigenfunction.
In the  case of a continuous spectrum, the label will usually be continuous or sometimes mixed, i.e. discrete+continuous.  
 Similarly, one can introduce two independent  $S$-wave eigenvectors:
\begin{equation} \label{eshsv}
 \mathbf{e}^{sh,sv}_{\mathbf{s}}( \mathbf{r})  = \frac{\bhat{p}^{sh,sv}}{\rho_0^{1/2} (2 \pi)^{3/2}} e^{i\mathbf{s}\cdot\mathbf{r}} ,
\end{equation}
where $\mathbf{\hat{p}}^{sh}$ is a unit vector perpendicular to $\bhat{s}$ and the vertical direction $\bhat{z}$, and
 $\mathbf{\hat{p}}^{sv}$ is a unit vector perpendicular to $\mathbf{s}$ and contained in the  $(\mathbf{z},\mathbf{s})$ plane.
The corresponding eigenvalue is $\beta_0^2 s^2$ where $\beta_0$ denotes the shear wave speed.
Note that the direction $\mathbf{z}$ is for the moment arbitrary but will be later taken perpendicular to the stratification.
The eigenvectors (\ref{ep})-(\ref{eshsv}) are orthogonal and  have the correct continuum normalisation in the sense of the scalar product 
(\ref{scprod}) defined
over $\field{R}^3$. Let us verify formally the orthogonality of say $ \mathbf{e}^{sh}_{\mathbf{s}}$ and $\mathbf{e}^p_{\mathbf{p}}$:
\begin{equation}
\begin{split}
 \langle  \mathbf{e}^{sh}_{\mathbf{s}} | \mathbf{e}^p_{\mathbf{p}} \rangle & = 
             \int\limits_{\field{R}^3}\!\! \rho_0 \frac{\bhat{p}^{sh} \cdot \bhat{p}^p   }{\rho_0 (2 \pi)^3}
                  e^{i\mathbf{r}\cdot\left(\mathbf{p} - \mathbf{s}\right)} d^3r \\
                & = \bhat{p}^{sh} \cdot \bhat{p}^p \delta\left(\mathbf{p} - \mathbf{s}   \right) .
\end{split}
\end{equation}
The delta function requires that $\mathbf{p} = \mathbf{s} $ in which case the scalar product of the polarization
 vectors vanishes. Thus the scalar product $ \langle  \mathbf{e}^{sh}_{\mathbf{s}} | \mathbf{e}^P_{\mathbf{p}} \rangle $ is identically
zero.  Let us now show formally that our set of eigenvectors is complete, i.e.:
\begin{equation}\label{I}
 \frac{\mathbf{I} \delta(\mathbf{r} - \mathbf{r'})}{\rho_0}  = \int\limits_{\field{R}^3}\!\! 
       \mathbf{e}^p_{\mathbf{p}}(\mathbf{r})  \mathbf{e}^p_{\mathbf{p}} (\mathbf{r'})^{\dagger} d^3p
    +  \int\limits_{\field{R}^3}\!\!   \mathbf{e}^{sh}_{\mathbf{s}}(\mathbf{r})  \mathbf{e}^{sh}_{\mathbf{s}} (\mathbf{r'})^{\dagger} d^3s
    +  \int\limits_{\field{R}^3}\!\! \mathbf{e}^{sv}_{\mathbf{s}'}(\mathbf{r})  \mathbf{e}^{sv}_{\mathbf{s}'}(\mathbf{r'})^{\dagger}  d^3s',
\end{equation}
where $ \mathbf{I} $ is the identity operator in polarization space:
$
   \mathbf{I}_{ij}  = \delta_{ij} $ and 
the symbol $^{\dagger}$ means hermitian conjugation (conjugate transpose).
  In equation (\ref{I}), the normalization prefactor  $\rho_0^{-1} $ is a consequence of the definition of the scalar product
 (\ref{scprod}).
  The sums in the right-hand side of equation (\ref{I}) can  be written as a single integral: 
\begin{multline}
  \int\limits_{\field{R}^3}\!\!   \mathbf{e}^p_{\mathbf{p}}(\mathbf{r})  \mathbf{e}^p_{\mathbf{p}} (\mathbf{r'})^{\dagger} d^3p
      +  \int\limits_{\field{R}^3}\!\!  \mathbf{e}^{sh}_{\mathbf{s}}(\mathbf{r})  \mathbf{e}^{sh}_{\mathbf{s}} (\mathbf{r'})^{\dagger} d^3s
             +  \int\limits_{\field{R}^3}\!\!  \mathbf{e}^{sv}_{\mathbf{s}'}(\mathbf{r})  
     \mathbf{e}^{sv}_{\mathbf{s}'}(\mathbf{r'})^{\dagger}  d^3s' = \\
      \frac{1}{(2 \pi)^3 \rho_0} \int\limits_{\field{R}^3} \!\!  \left( \bhat{p}^p \bhat{p}^{p^{\dagger}} + 
        \bhat{p}^{sh}  \bhat{p}^{sh^{\dagger}} 
                  +   \bhat{p}^{sv} \bhat{p}^{sv^{ \dagger}}  \right)  
                e^{i \mathbf{p} \cdot (\mathbf{r} - \mathbf{r}') }d^3p  = \\
         \frac{ \mathbf{I}}{2 \pi^3 \rho_0}      \int\limits_{\field{R}^3}\!\!       e^{i \mathbf{p} \cdot (\mathbf{r} - \mathbf{r}') }d^3p = 
          \frac{ \mathbf{I} \delta(\mathbf{r} - \mathbf{r}') }{\rho_0} ,
\end{multline}
because  the set $(\bhat{p}^p,\bhat{p}^{sh},\bhat{p}^{sv})$ forms an orthonormal basis in polarization   space.
We have therefore verified the formal completeness relation.
\subsection{Eigenvectors of the elastic wave equation in a stratified half-space}
To construct a complete set of eigenvectors in a stratified Earth, we will apply scattering theory. One key idea
which can be found e.g. in  \cite{reed79} is the following. Let us assume that a complete
set of eigenfunctions $\mathbf{e}^0_k $  of the wave operator $\mathbf{A_0}$  in a reference medium is known. Now, perturb the
reference medium by introducing a localized  scattering body to obtain a new operator $\mathbf{A}$.
 Then it is possible to show that  the eigenvectors 
 $  \mathbf{e}_k $  of the perturbed operator  are obtained by considering the scattering of 
$ \mathbf{e}^0_k $ by the heterogeneous body. More precisely,  
$ \mathbf{e}_k $  is the solution of the Lippman-Schwinger equation:
\begin{equation}  \label{LSE}
 \mathbf{e}_k  =   \mathbf{e}^0_k  +      \mathbf{ G^0 A'} \mathbf{e}_k ,
\end{equation}
where $ \mathbf{G^0}$ is the retarded Green function associated with the operator $\mathbf{A_0}$. The second term
on the right-hand side of equation (\ref{LSE}) represents the scattered wave which results from the interaction
between an unperturbed incoming wave and the scatterer represented by $\mathbf{A'}$, a perturbation of  $\mathbf{A_0}$.
 The set of eigenvectors constructed 
in this way is orthogonal and properly normalized (but not necessarily complete). 
 Similar results have been proved for wave scattering by infinite boundaries by \citet{ramm86}.
The construction of a complete basis of eigenvectors for fluids with arbitrary stratification has been conducted by
\citet{wilcox84}. This author demonstrates how the complete set of eigenfunctions can be constructed from the scattering of plane
waves incoming from infinity. The method of Wilcox has been used by \citet{dermenjian85} and \citet{secher98} to construct 
rigorously a complete basis of eigenfunctions of the elastic operator in a homogeneous 
 half-space with a traction free boundary. Applying the
same technique, we can exhibit a complete set of eigenvectors for a stratified half-space. 
Complete justification of the formulas below are presented in a separate  paper \citep{margerin08}.
For an alternative approach the reader is referred to \citet{maupin96}.
We assume that as $z \to -\infty$,
the density and Lam\'e parameters tend to $\rho_{\infty}$, $\lambda_{\infty}$ and $\mu_{\infty}$, respectively. 
The decoupling between $SH$ and $(P,SV)$ motions defines two families of solutions.
The case of $SH$ waves is completely similar to the acoustic case described by \citet{wilcox84} and will not be discussed. 
For the $P-SV$  case we take advantage of the translational invariance of the problem and 
look for solutions of the wave equation which have the following form as $z \to -\infty$:
\begin{equation} \label{modep}
  \mathbf{\psi}^p_{\mathbf{p}}( \mathbf{r} )  \to   \frac{1}{(2 \pi)^{3/2} \rho_{\infty}^{1/2}} 
                                               \left( \bhat{p} e^{i\mathbf{p} \cdot \mathbf{r}} +  r^{pp} \bhat{p}^r e^{i\mathbf{p}^r \cdot \mathbf{r}}
                                                    + r^{ps} \bhat{p}^{sv} e^{i\mathbf{s}\cdot\mathbf{r}} \right),
\end{equation}
where the wavevectors $\mathbf{p}$, $\mathbf{p}^r  $, and $\mathbf{s} $ have the following properties:
\begin{align}
\mathbf{p} & = (p_x,p_y,p_z) , \quad (p_x,p_y) \in \field{R}^2  \text{,  } p_z \in \field{R}^+, \\
\mathbf{p}^r & = (p_x,p_y,-p_z) , \\
\mathbf{s} & = \left(p_x,p_y,-\sqrt{p_{\parallel}^2\left(\alpha^2_{\infty}/\beta^2_{\infty} -1 \right) +p_z^2 \alpha^2_{\infty}/\beta^2_{\infty}}\right) 
                  , \quad  p_{\parallel}^2 =p_x^2 + p_y^2 .
\end{align} 
As before,  $\bhat{p}^r$ is a unit vector in the direction of $\mathbf{p}^r$ and 
$\bhat{p}^{sv}$ is a unit vector orthogonal to $\mathbf{s}$, contained in the $(\bhat{z},\mathbf{s})$ plane. 
The $P$ and $S$ wavespeeds
at infinity are denoted by $\alpha_{\infty}$ and $\beta_{\infty}$, respectively. In equation (\ref{modep}), the dependence of the eigenvector
on the wavenumber of the incident wave is emphasized. This equation can be understood as follows: a generalized eigenvector of the operator
 $\mathbf{A}$ is the sum of an eigenvector of  $\mathbf{A_0}$, plus the wave back-scattered by the layered structure. The reflection
coefficients are determined by the continuity of traction and displacements across interfaces, and by the vanishing of tractions at the
free surface. Methods to calculate the reflection matrix  $R$ of the layer stack  
are described in e.g. \citet{ken83}  and \citet{aki02}.
Equation (\ref{modep}) defines eigenvectors of $\mathbf{A}$ with eigenvalue $\omega^2 = \alpha_{\infty}^2 p^2$.
Another family of solution corresponds to incident $SV$ waves from infinity $(z \to -\infty)$:
\begin{equation} \label{modesv}
  \mathbf{\psi}^s_{\mathbf{s}}(\mathbf{r})  \to  
   \frac{1}{(2 \pi)^{3/2} \rho_{\infty}^{1/2}} 
                                               \left( \bhat{p}^{sv} e^{i\mathbf{s}\cdot\mathbf{r}} + r^{sp} \bhat{p} e^{i\mathbf{p}\cdot\mathbf{r}}
                                                    + r^{svsv} \bhat{p}^{sv_r} e^{i\mathbf{s}^r \cdot \mathbf{r}} \right), 
\end{equation}
where the  wavevector $\mathbf{p}$   must be carefully defined, depending on $\mathbf{s} = (s_x,s_y,s_z)$, $s_z >0$: 
\begin{equation}
 \mathbf{p}  = 
 \begin{cases}
    \left(s_x,s_y,\sqrt{s_{\parallel}^2\left(\beta^2_{\infty}/\alpha^2_{\infty} -1 \right) + s_z^2 \beta^2_{\infty}/\alpha^2_{\infty}}  \right)
                  & \quad s_z > s_{\parallel} \sqrt{ \frac{\alpha_{\infty}^2}{\beta_{\infty}^2} -1  } \\
      \left(s_x,s_y,-i\sqrt{s_{\parallel}^2\left(1 - \beta^2_{\infty}/\alpha^2_{\infty}  \right) - s_z^2 \beta^2_{\infty}/\alpha^2_{\infty}}  \right)
                & \quad  0< s_z < s_{\parallel} \sqrt{ \frac{\alpha_{\infty}^2}{\beta_{\infty}^2} -1}
 \end{cases} ,
\end{equation}
where $s_{\parallel}^2 = s_x^2 + s_y^2 $.
The eigenvectors (\ref{modesv}) have eigenvalues $\omega^2 = \beta_{\infty}^2 s^2$.
 Note that beyond the critical angle for $SV-P$ mode conversion 
-- $0<s_z< s_{\parallel} \sqrt{ \frac{\alpha_{\infty}^2}{\beta_{\infty}^2} -1} $--,
the reflection coefficients as well as the polarization vector $\mathbf{\hat{p}}$ become complex.
 The reflected $P$ wave is evanescent in the lower half-space, and decays exponentially with depth.
The complex polarization vector carries information on the particle motion of  $P$ waves in the half-space.
Since the vertical component has a $-\pi/2$ phase shift with respect to the horizontal component,
the evanescent $P$ wave is in  prograde elliptical motion.
Equations (\ref{modep})-(\ref{modesv}) define the generalized eigenfunctions that have oscillatory behavior at infinity.
However, they do not form a complete set. Similar to bound states that decay rapidly away from a scattering body, the set of eigenvectors
must be complemented with  surface waves, that have their energy localized near the traction-free surface. 
In the case of surface waves, the set of eigenvalues is continuous but organized
along discrete dispersion branches. The dispersion relation of the surface wave modes can be determined
by finding the poles of the reflection coefficients located on the positive real axis. Numerical procedures are described in the books
of \citet{ken83} and \citet{aki02}.  The Rayleigh wave eigenfunctions have the form:
\begin{equation}
\label{eq:rayl}
    \mathbf{\psi}^{Rl}_{\mathbf{p_{\parallel}},n}(\mathbf{r})  = \frac{1}{2 \pi} 
  \left(\frac{p_x \phi^x_{(\mathbf{p}_{\parallel},n)}(z)}{\sqrt{p_x^2 + p_y^2}},\frac{p_y  \phi^x_{(\mathbf{p}_{\parallel},n)}(z)}{\sqrt{p_x^2 + p_y^2}},
\phi^z_{(\mathbf{p}_{\parallel},n)}(z) \right)^t
      e^{i (p_x x + p_y y )}, \quad  n \in \field{N}
\end{equation} 
where  $ \phi^x_{(\mathbf{p}_{\parallel},n)}$ and $\phi^z_{(\mathbf{p}_{\parallel},n)}$ denote the horizontal and vertical components  of the n$^{th}$ Rayleigh eigenfunction, 
and obey the following normalization relation: 
\begin{equation}
  \label{eq:normrayl}
  \int_{\field{R}^-}\!\! dz \rho(z) 
 \left( \left| \phi^x_{(\mathbf{p}_{\parallel},n)}(z) \right|^2 + \left| \phi^z_{(\mathbf{p}_{\parallel},n)}(z) \right|^2\right)  =1 . 
\end{equation}
 These eigenvectors have eigenvalues $\omega_{n}^2 = c_n(p_{\parallel})^2 p_{\parallel}^2$, where $c_n$ denotes
the phase speed of mode $n$.  Note that in equation (\ref{eq:rayl}),
the surface wave  eigenvector is labeled simultaneously with the continuous  indices $\mathbf{p}_{\parallel}= (p_x,p_y)$ and the discrete mode
branch index $n$. This means that if we fix the horizontal wavevector $\mathbf{p}_{\parallel}$ and look for solutions of the 
wave equation that verify simultaneously the traction-free condition and the vanishing of traction and displacements
at depth, this can only occur at a discrete set of eigenfrequencies denoted by  $\omega_1, \cdots, \omega_n, \cdots $.
Note that in applications, we will work the other way around:  
the central frequency of the signal is fixed and one looks for poles of the reflection coefficients 
related to plane waves with apparent velocities smaller than the shear wave velocity in the lower half-space.
When a candidate mode has been identified,  the Rayleigh quotient is used to calculate 
 its group velocity and as an accuracy check. 
In this formulation, it is implicit that the n$^{th}$ mode appears beyond the cut-off frequency $\omega_n^c$.
 \section{Equipartition Theory}
\subsection{Formal definition of equipartition}
Following  \citet{shapiro00},  we introduce the compression and shear deformation energies:
\begin{align}
\label{eq:ecomp}
W^{p}  = &\frac{1}{2}\rho_0\alpha_0^2(\nabla \cdot \mathbf{u})^2 \\
\label{eq:eshear}
W^{s}  = &\frac{1}{2}\rho_0\beta_0^2(\nabla \times \mathbf{u})^2
\end{align}
where $\alpha_0$, $\beta_0$ and $\rho_0$ denote the compressional, shear velocity and the density at the receiver, 
and $\mathbf{u}$ is the displacement vector.
\citet{weaver82,weaver90} and \citet{ryzhik96} showed that
in a heterogeneous infinite elastic medium, the {\it P} to {\it S} energy ratio stabilizes at large lapse time:
\begin{equation}
\lim_{t \rightarrow +\infty} \frac{\langle W^{s}\rangle }{\langle W^{p}\rangle } =
\frac{2\alpha_0^3}{\beta_0^3} ,
\label{limite_infinie}
\end{equation}
where $\langle .\rangle $ is a statistical averaging over the configuration of
scatterers in the medium.
This result means that in a heterogeneous medium,
multiple scattering creates a partition of compressional and shear energies, which is
reached after the waves have encountered a sufficiently large number of
scatterers. This is the physical interpretation of the limit $t \rightarrow +\infty$. 
Remarkably, the equilibration ratio is completely independent of the nature of the scatterer
and of the source type.

The result (\ref{limite_infinie}) gives no information about the time needed to reach
equilibrium. The significance of the limit $t \rightarrow +\infty$
was addressed theoretically by \citet{margerin00} and experimentally by \cite{malcolm04} and \citet{paul05}. By solving numerically the elastic 
radiative transfer equation, \citet{margerin00} showed that the asymptotic behavior occurs after a few mean 
free times only, where  the mean  free time denotes the average time between two scattering events.
These authors also showed that the equilibration time depends on the type of scatterers
 and the type of source. Therefore, the dynamics of the stabilization process contains
information on the medium heterogeneities.
It is also important to realize that equilibration is different from equipartition. Equipartition demands
that all the modes be represented with equal energies which in turn implies that the flow of
energy is isotropic, at least away from medium boundaries \citep{malcolm04}. In a recent study, \citet{paul05} have shown that equipartition
implies equilibration but that the converse statement is wrong. In particular,  Equation (\ref{limite_infinie}) applies 
even when the energy flux distribution  is still strongly anisotropic.

It is mathematically convenient to define equipartition as a white noise distributed
over all the modes of the system \citep{weaver82,hennino01}. In order to illustrate this definition,
let us derive the $W^s/W^p$ ratio at equipartition in an infinite homogeneous 3-D medium using the  eigenvectors  
 (\ref{ep})-(\ref{eshsv}). Completeness implies that the complex vector field $\mathbf{u}$ can be expanded
as follows:
\begin{equation}
  \label{eq:fieldu}
  \mathbf{u}\left(t,\mathbf{r} \right) = \int\limits_{\field{R}^3}\!\!  \mathbf{e}_{\mathbf{p}}^p(\mathbf{r} )   e^{-i \omega_\mathbf{p}t} d^3p
                                         + \int\limits_{\field{R}^3}\!\! \left[ a_{\mathbf{s}}^{sh}  \mathbf{e}_{\mathbf{s}}^{sh}(\mathbf{r})   
                                         + a_{\mathbf{s}}^{sv}   \mathbf{e}_{\mathbf{s}}^{sv}(\mathbf{r} ) \right]  e^{-i \omega_\mathbf{s}t} d^3s
\end{equation}
where $\omega_p = \alpha_0 \sqrt{p_x^2 + p_y^2 + p_z^2}$,  $\omega_s = \beta_0 \sqrt{s_x^2 + s_y^2 + s_z^2}$.
 Equation (\ref{eq:fieldu}) defines the analytic signal associated with the measured displacements.
At equipartition, the amplitudes $a_{\mathbf{p}}^p$, ... are uncorrelated random variables with zero mean and equal variance. 
In practice,  they   are slowly varying functions of time in the frequency band of interest. In order to investigate the energy
content at time $t$ in the coda, it is convenient to introduce the Wigner distribution of the wavefield:
\begin{equation}
  \label{eq:wigner}
  E_s(t,\tau, \mathbf{r}) = \frac12 \rho_0 \beta_0^2 \left\langle \nabla\times \mathbf{u}(t+\tau/2, \mathbf{r}) 
   \cdot \nabla\times \mathbf{u}(t-\tau/2, \mathbf{r})^* \right\rangle,
\end{equation}
where $^*$ denotes complex conjugation. In equation (\ref{eq:wigner}), the brackets denote an ensemble average.
Assuming that the coefficients $a_{\mathbf{s}}$,  $a_{\mathbf{p}}$ are governed by a quasi-stationary white-noise  process:
 \begin{equation}
  \label{whitenoise}
 \langle a_{ \mathbf{s}}(t-\tau/2) a^*_{ \mathbf{s'}}(t+\tau/2)  \rangle = \sigma^2(t) \delta( \mathbf{s}- \mathbf{s'}),
 \end{equation}
one obtains:
\begin{equation}
  \label{eq:wigner2}
  E_s(t,\tau, \mathbf{r}) = \frac12 \rho_0 \beta_0^2 \sigma^2(t) \sum_{m= {sh,sv}} \int\limits_{\field{R}^3}\!\!
       \left| \nabla\times \mathbf{e}^{m}_{\mathbf{s}}(\mathbf{r})   \right|^2 e^{-i \omega_{\mathbf{s}} \tau} d^3s.
\end{equation}
In seismological applications, the function $\sigma$  models the slow decay of the energy envelope in the coda.
The Fourier transform of the Wigner distribution gives the power spectral density at frequency $\omega_0$:
\begin{eqnarray}
E_s(t,\omega_0,\mathbf{r}) & = &   \frac{1}{2 \pi} \int\limits_{-\infty}^{+\infty}\!\!  E_s(t,\tau, \mathbf{r})  e^{i\omega_0 \tau }d\tau \\
\label{eq:wigner3}
 & = & \frac12 \rho_0 \beta_0^2 \sigma^2(t)
          \sum_{m=sh,sv} \int\limits_{\field{R}^3}\!\! \left| \nabla\times   \mathbf{e}^m_{\mathbf{s}} (\mathbf{r})  \right|^2  
        \delta(\omega_0 -\omega_{\mathbf{s}}) d^3s .
\end{eqnarray}
The assumption of quasi-stationarity demands that the function $\sigma$ does not vary significantly during
one cycle $2 \pi/\omega_0$, where $\omega_0$ is the central frequency of the analyzed signal.
Using equation (\ref{eshsv}), one can show that the squared modulus
of the rotational part of the eigenvectors of a homogeneous full-space is independent of position:
 \begin{equation}
 \left| \nabla\times  \mathbf{e}^{sh}_{\mathbf{s}}( \mathbf{r})  \right|^2 = \frac{\omega_0^2}{\rho_0 \beta_0^2 (2 \pi)^3}.
 \end{equation}
Introducing spherical polar coordinates to  perform the wave number integral in equation (\ref{eq:wigner3}),  
 one finds the  total shear energy spectral density:
\begin{equation}
E_s(t,\omega_0) =\frac{ 2 \sigma^2(t) \omega_0^4}{(2 \pi)^2\beta_0^3}.
\end{equation}
Using the definition of the compressional energy (\ref{eq:ecomp})  and following similar lines, one obtains
the total compressional  energy spectral density:
\begin{equation}
E_p(t,\omega_0) =  \frac{  \sigma^2(t) \omega_0^4}{(2 \pi)^2\alpha_0^3},
\end{equation}
from which  the result (\ref{limite_infinie}) follows.

\subsection{Equipartition in a stratified half-space}
In the case of a stratified half-space, calculations have to be performed numerically and in general, the
energy ratios depend on  depth. Near the boundaries the coupling between
{\it P} and $S$ waves complicates the physical interpretation of the  decomposition of the field energy into shear and compressional
components. For instance, \citet{shapiro00}  have shown that, depending
on the incidence angle at the free surface of a homogeneous half-space, a shear wave can generate a large
amount of compressional  energy and vice-versa. In addition, the sum of
 compressional and shear energies does not equal the total elastic deformation energy.
However, the ratio of shear to compressional energy, as defined above, constitutes
 a  marker of the wave content of the seismic coda. 
We will also consider the vertical to horizontal kinetic energy ratio.
This quantity can be measured without major difficulties on a rather broad
frequency range and offers additional information.
To illustrate the computational approach, let us for instance consider the contribution 
of the $n^{th}$  Rayleigh mode to the vertical kinetic energy $E^{Rl}_z$.
Using equation (\ref{eq:rayl}) and the property (\ref{whitenoise}), one finds:
\begin{equation}
  \label{eq:ecz1}
  E^{Rl}_z(t,\omega_0,z) =\frac{\rho(z)\omega_0^2 \sigma^2(t)}{8\pi^2}  \int\limits_{\field{R}^2}\!\! \left| \phi^z_{(\mathbf{p}_{\parallel},n)}(z) \right|^2 
    \delta \left(\omega_0 - \omega_{\mathbf{p}_{\parallel}} \right) d^2p_{\parallel},  
\end{equation}
 where $ \omega_{\mathbf{p}_{\parallel}} = c_n(\mathbf{p}_{\parallel}) p_{\parallel}  $,
and the vector $\mathbf{p}_{\parallel}$ has components $(p_x,p_y,0)$. 
Introducing cylindrical coordinates, the wavenumber integral can be performed to obtain:
 \begin{equation}
  \label{eq:ecz2}
  E^{Rl}_z(t,\omega_0,z) = \frac{\rho(z) \omega_0^3 \sigma^2(t)}{c_n u_n 4 \pi} \left| \phi^z_{(p_{\parallel} =\omega_0/c_n,n)}(z) \right|^2 ,
\end{equation}
where $u_n$ denotes the group velocity of mode $n$ at frequency $\omega_0$. The transition from equation
 (\ref{eq:ecz1}) to (\ref{eq:ecz2})  uses the following  decomposition of the delta function:
\begin{equation}
\delta\left( \omega_0 -  \omega_{\mathbf{p}_{\parallel}} \right) = \frac{1}{\left| u_n  \right|}
   \delta\left(p_{\parallel} - \frac{\omega_0}{c_n}\right). 
\end{equation}
Equation (\ref{eq:ecz2}) agrees with results of \cite{tregoures02} who developed a transport
theory for elastic waves in a plate.
The interpretation of equation (\ref{eq:ecz2}) is as follows: in the seismic coda, the contribution of
the $n^{th}$ Rayleigh mode to the vertical kinetic energy is proportional to the global density of states of the
mode and the local value of the  wavefunction squared. This product is often referred to as the ``local density of states''
in the literature \citep{economou05}.

\begin{figure}
\vspace*{1cm}
\begin{center}
\epsfxsize=12cm
\epsfbox{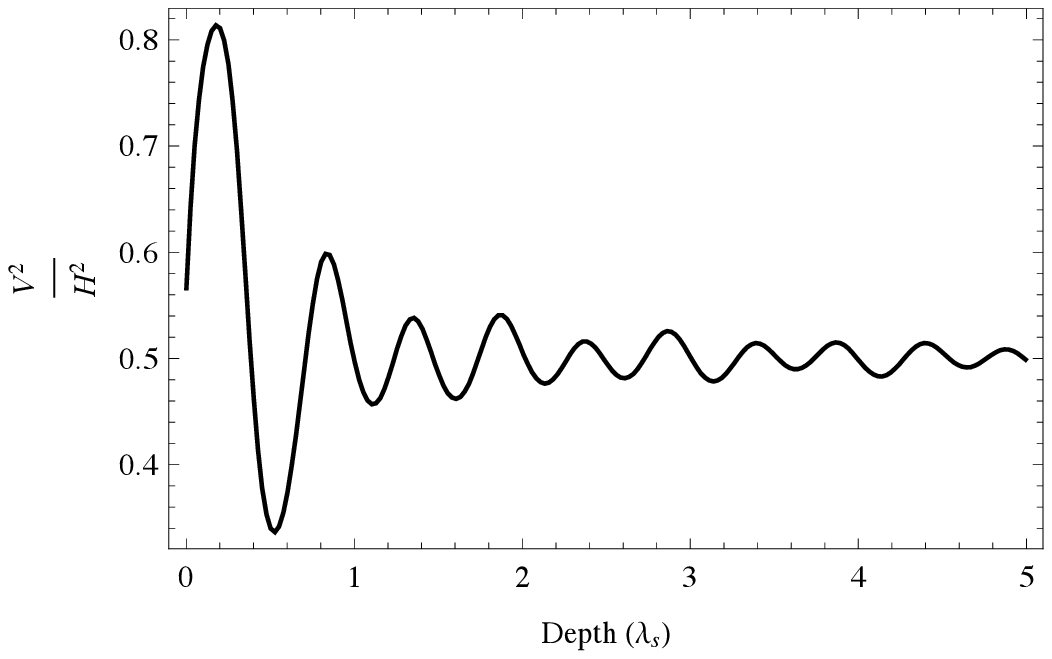}

\vspace*{0.4cm}
\epsfxsize=12cm
\epsfbox{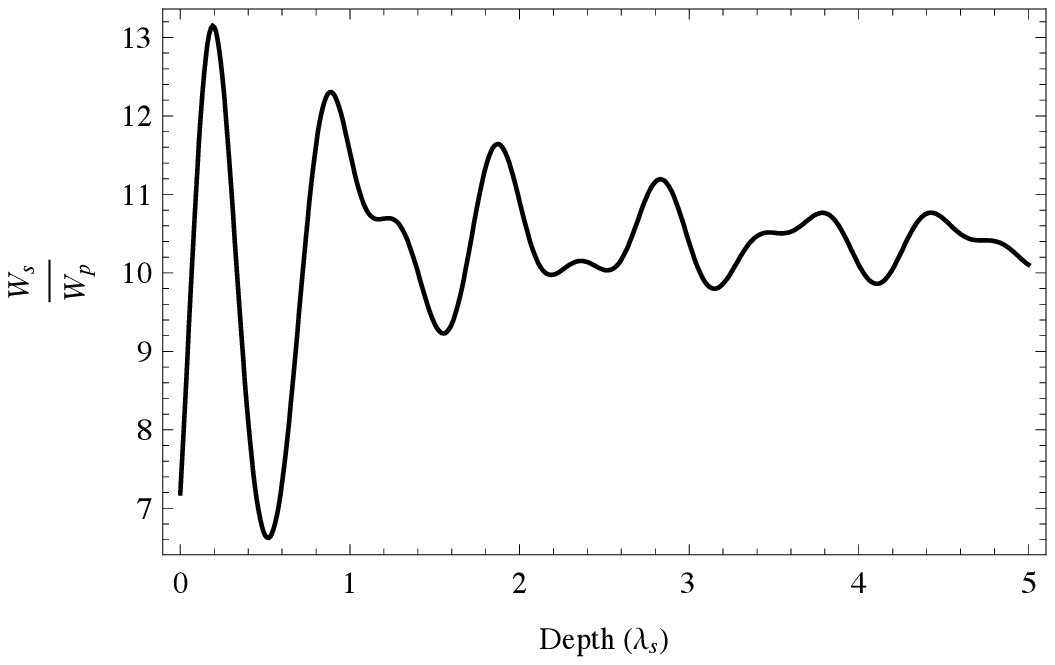}
\end{center}
\caption{Depth dependence of energy ratios at equipartition in a homogeneous elastic half-space (Poisson solid). 
The depth unit is the shear wavelength $\lambda_s$.  Top: vertical to horizontal kinetic energy ratio.
Bottom: shear to compressional deformation energy ratio. Note the persistent oscillations at depth, which originate
 from the interference between incident and reflected waves at the free surface.}
\label{halfspace}
\end{figure}

It is also instructive to consider the expression of the  kinetic energy for the generalized eigenfunctions
 representing body waves in the lower half-space. For simplicity, we consider the
 kinetic energy of vertical vibrations $E^p_z$  for the generalized eigenfunctions (\ref{modep}) corresponding to $P$ waves incident from below
the sub-surface structure, together with their reflections. To facilitate the comparison with the case of Rayleigh waves,
we can rewrite the generalized $P$ wave eigenvector as follows: 
\begin{equation}
\label{psip}
 \psi^p_{(\mathbf{p_{\parallel}},p_z)}(\mathbf{r}) =  \frac{e^{i p_x x + i p_y y}}{2 \pi} \phi^p_{(\mathbf{p_{\parallel}},p_z)}(z), 
\end{equation}
i.e. we split  the  horizontally and vertically propagating parts. The decomposition (\ref{psip}) highlights the similar roles
played by  the  surface wave mode index $n$ and the   vertical wavenumber $p_z$.
 Using equations (\ref{modep}) and (\ref{whitenoise}),  we obtain the following expression:
\begin{equation}
  \label{eq:epz1}
    E^{p}_z(t,\omega_0,z) =  \frac{\rho(z) \omega_0^2 \sigma^2(t)}{8 \pi^2} \int\limits^{+\infty}_{0} dp_z  \int\limits_{\field{R}^2}\! 
            \left| \phi^z_{(\mathbf{p}_{\parallel},p_z)}(z) \right|^2 
    \delta \left(\omega_0 - \omega_{(\mathbf{p}_{\parallel},p_z)} \right) dp_{\parallel},
\end{equation}
where $ \omega_{(\mathbf{p}_{\parallel},p_z)} = \alpha_{\infty} \sqrt{p_x^2 + p_y^2 + p_z^2}, $ and $\phi^z_{(\mathbf{p}_{\parallel},p_z)}$
denotes the vertical component of the generalized $P$-wave eigenfunction, which depends explicitly on the wavenumbers
 $\mathbf{p}_{\parallel}$ and $p_z$. The main difference between equations (\ref{eq:epz1}) and (\ref{eq:ecz1}) is the substitution
of the discrete mode branch $n$ with the continuous vertical wavenumber $p_z$.
After introducing cylindrical coordinates,
 integrating over the vertical wavenumber and the azimuthal angle, one obtains:
\begin{equation}
  \label{eq:epz2}
    E^{P}_z(t,\omega_0,z) = \frac{\rho(z) \omega_0^3 \sigma^2(t)}{4 \pi \alpha_{\infty}^2} 
    \int\limits_0^{\omega_0/\alpha_{\infty}} \!\! dp_{\parallel} 
          \frac{ p_{\parallel}}{ \sqrt{\frac{\omega_0^2}{\alpha_{\infty}^2} -p_{\parallel}^2}} 
       \left| \phi^z_{(\mathbf{p}_{\parallel},p_z )}(z)
  \right|^2 \bigg\vert_{p_z= \sqrt{\frac{\omega_0^2}{\alpha_{\infty}^2} -p_{\parallel}^2}} ,
\end{equation}
where $\alpha_{\infty}$ is the $P$ wave speed in the underlying half-space.
Upon making the change of variable $p_{\parallel}=\frac{\omega_0}{\alpha_{\infty}}\sin \theta $, the last equation simplifies
to:
\begin{equation}
  \label{eq:epz3}
   E^{P}_z(t,\omega_0,z) =  \frac{\rho(z) \omega_0^4 \sigma^2(t)}{4 \pi \alpha_{\infty}^3} 
     \int\limits_0^{\pi/2} \!\! d\theta 
    \sin \theta   \left| \phi^z_{(\mathbf{p}_{\parallel},p_z )}(z)
  \right|^2 \bigg\vert_{\left( p_{\parallel}=\frac{\omega_0}{\alpha_{\infty}} \sin \theta,p_z= \frac{\omega_0}{\alpha_{\infty}} \cos \theta \right)}.
\end{equation}
Evaluated at $z=0$ in the case of a homogeneous half-space,  equation (\ref{eq:epz3}) reduces to similar
 expressions obtained  by \citet{weaver85}, based on physical reasoning. 
 Note that away from the free surface, the eigenmodes of the system do not reduce to pure
$P$ or $S$ waves but consist of a superposition of incident and reflected body waves.
The result (\ref{eq:epz3}) can be interpreted as follows: the contribution of generalized
 $P$ eigenfunctions to the kinetic energy of vertical vibrations 
 is proportional to the density of states of  $P$ waves  in the lower half-space and to the
  squared $P$-wavefunction, averaged over all possible directions of $P$-waves incident from below the structure. Note that
the last integral (\ref{eq:epz3}) also depends on  depth and, in general, presents complicated oscillations.  
This is illustrated in Figure \ref{halfspace} where kinetic and deformation energy ratios are plotted
as a function of depth in a homogeneous 3-D Poisson half-space. 
The largest fluctuations of the deformation and kinetic energy ratios occur in the vicinity of the free surface and are mostly caused by
the depth dependence of the Rayleigh wave eigenfunction.
When averaged over one shear  wavelength $\lambda_s$ at a depth greater than 3 $\lambda_s$,
the energy ratios reduce to the expected values for a homogeneous infinite medium.
Our calculations are in perfect agreement with previous work by \citet{hennino01,tregoures02}, based 
on a locked-mode method. The present approach is  generally valid for an arbitrarily stratified half-space.   
Experimentally, the $V^2/H^2$ kinetic energy ratio  is fairly easy to measure over a broad frequency range but can be strongly affected by the
 local topography. The ratio of deformation energies $W^s/W^p$ is more difficult to evaluate but is independent of the choice
of local coordinate system. It is therefore expected to be less sensitive to geometrical effects.
In the next section, we discuss the measurement of the deformation energy ratio on a 
dense array.

\section{Stabilization of deformation energy ratios in the seismic coda}

\subsection{Data set and pre-processing}

\begin{figure}
\begin{center}
\epsfxsize=12cm
\epsfbox{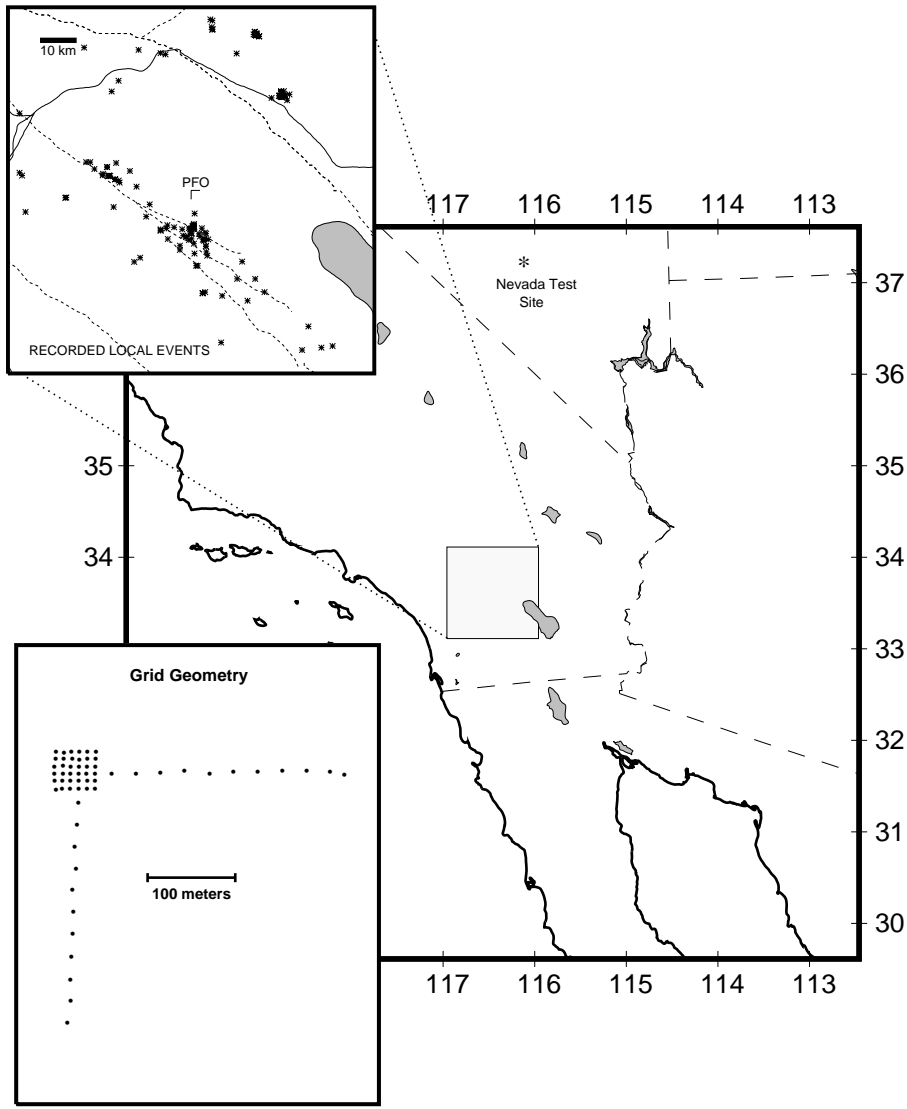}
\end{center}
\caption{Location map of the 1990 Pinyon Flats High Frequency Array Experiment. 
58  3-components sensors were deployed during 3 months. The upper-left inset  shows the location of
the  largest magnitude local events 
recorded by the array. The lower-left inset shows the acquisition geometry: a dense square of 
$6 \times 6$ seismometers separated by 7 meters,   and two perpendicular long arms 
composed of  11 seismometers separated by 21 meters. Reproduced from \citet{vernon98}}
\label{fig:carte}
\end{figure}

The data we use were collected by a temporary array that was deployed at the Pinyon Flats
Observatory (PFO) in
1990. This array operated as part of the Incorporated Research
Institutions for Seismology (IRIS), and was located in Southern 
California between the San Jacinto
and the San Andreas fault, as shown on the location maps in Figure~\ref{fig:carte}

The array was composed of 58 sensors arranged in a six-sensor by
six-sensor grid with 200m long arms. The sensors were 2Hz
L22-D geophones with an acquisition sampling frequency of 250Hz.
 They were triggered by two borehole sensors located at 275m and 
150m depth, respectively.  A total of 300 events was recorded, among which 140
 were located with good accuracy. In our study, we have used 10 events whose source
parameters are described in Table~\ref{sourceparam}. These earthquakes have  
epicentral distances smaller than 50km and magnitudes higher than 2. 
We chose these events because they exhibit  a pronounced coda with
 a high signal to noise ratio.
We took care of filtering out the frequency components above  40Hz because the
signal exhibits significant instrumental noise around 60Hz.
Because the estimate of the divergence and curl of the wavefield
requires the application of the traction-free boundary condition \citep{shapiro00}, we must first  
define geometrically the free surface. For this purpose we chose the plane which
best approximates  the location of stations in a least-squares  sense.
 The data were   rotated in a new reference system whose vertical axis is directed perpendicular
 to the optimal plane.
The signal components  were detrended and normalized to equal root mean square amplitude at each station $i$:
 \begin{equation}\label{normalisation}
 \forall i \in \left\{ 1,\cdots,n \right\}, \quad\begin{cases}
  {\displaystyle \int \!\!  u_z(\mathbf{r}_i,t)^2 dt =\overline{u}_z^2} \\ 
 {\displaystyle   \int \!\!  u_{ew}(\mathbf{r}_i,t)^2 dt =\overline{u}_{ew}^2} \\ 
  {\displaystyle \int \!\!   u_{ns}(\mathbf{r}_i,t)^2 dt =\overline{u}_{ns}^2 }
 \end{cases}   ,
 \end{equation}
where the bar denotes the root mean square.
This normalisation procedure  corrects for possible
instrumental response differences and yields more reliable estimates for the field derivative.

\subsection{Method of estimation for $P$ and $S$ energies}
\begin{figure}
\begin{center}
\epsfxsize=16cm
\epsfbox{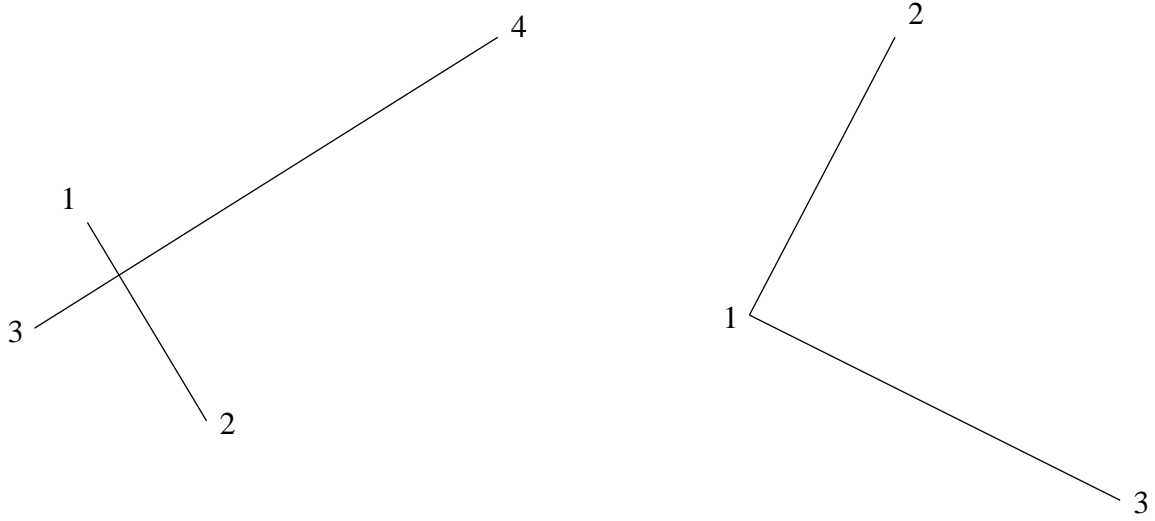}
\end{center}
\caption{Estimation of the divergence and curl of the wavefield requires the measurement of  the derivative along two linearly
 independent directions. This figure presents two valid configurations with 3 or 4 stations}
\label{config}
\end{figure}

\begin{figure}
\begin{center}
\includegraphics[width=0.8\linewidth]{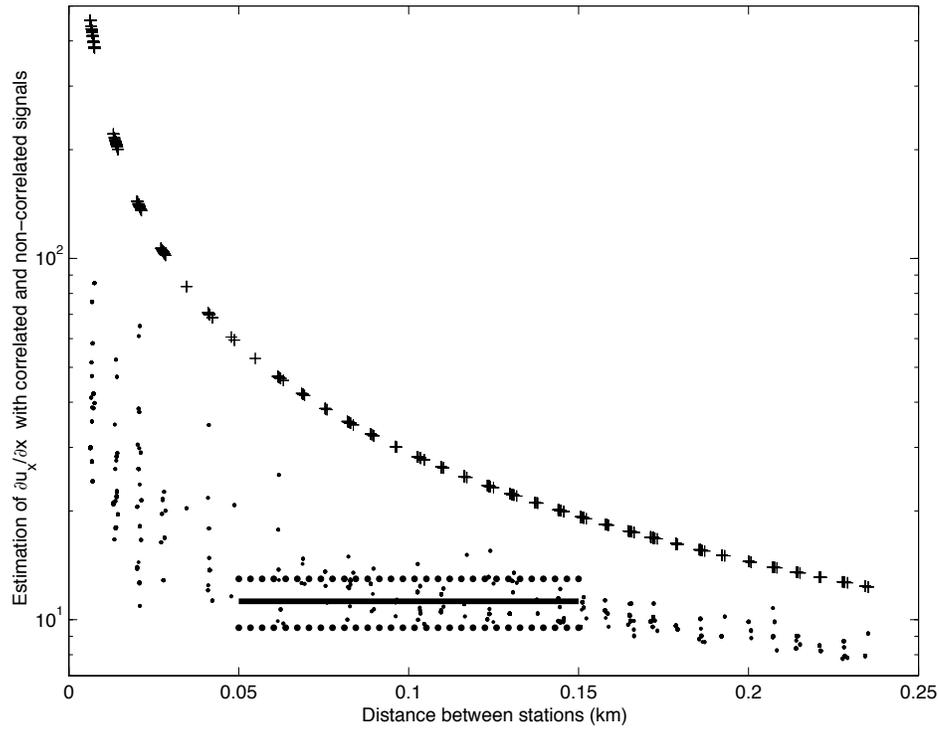}
\end{center}
\caption{ Measurement of the spatial derivative of the wavefield on a dense array of seismometers.
 Dots: estimates of the derivative
  $\partial{u_z}/\partial{x}$ for one event at lapse time $t = 32.4$s. 
 The  mean value of the field derivative is indicated by a solid line. 
The horizontal dotted lines indicate the plus or minus one standard deviation range.
The $+$ symbols show the estimate  of the derivative for incoherent fields.
The plot illustrates the  stability of the estimate of the derivative
 for  stations located at least 50m and at  most 150m  apart. Beyond this distance the field at the two stations
 become uncorrelated as shown by the coincidence between the incoherent and coherent estimates of the wavefield 
derivative.}
\label{stab_der}
\end{figure}

To estimate the deformation energies $W^p$ and $W^s$,  the displacement gradients
 $\partial_iu_j$ of the wavefield need to be measured. 
We will first examine the problem
of estimating the horizontal derivatives. These will be subsequently  used to obtain the vertical derivatives.
 Using three or four stations as shown in Figure \ref{config}, the derivative of the wavefield in two linearly
independent horizontal directions can be estimated, which suffices to recover the horizontal
 components of the gradient vector by application of the Taylor expansion formula:
\begin{equation}
\label{eq:taylor}
u(\mathbf{r}_2)\approx u(\mathbf{r}_1) + (\mathbf{r}_2-\mathbf{r}_1).\mathbf{\nabla}u +
   O \left(\left| \mathbf{r}_2-\mathbf{r}_1 \right|^2 \right) .
\end{equation}
In practice, the  estimate of the spatial derivative requires a great coherence
of the displacements at two near-by stations. 
Applying a finite difference approximation to equation (\ref{eq:taylor}) , one finds
\begin{equation}
\label{approx_der}
\frac{\partial{u_i}}{\partial
  x_j}\approx\frac{u_i(\mathbf{r}_1+d.\mathbf{j})-u_i(\mathbf{r}_1)}{d},
\end{equation}
where $d$ is the inter-station distance. 
 $u_i(\mathbf{r}_1+d.\mathbf{j})$ and $u_i(\mathbf{r}_1)$ correspond to  the 
measured displacements at positions $\mathbf{r}_2=\mathbf{r}_1+d \mathbf{j}$ and $\mathbf{r}_1$, respectively.
The approximation (\ref{approx_der}) will be correct provided $d/\lambda\ll 1$ and its application
to the data set will be critically examined.
Following  \citet{shapiro00}, we use the free surface boundary condition to express the
derivative in the $z$ direction in terms of horizontal derivatives:
\begin{equation}
 \left\{ \begin{array}{l}
               \partial_z{u_x}  =- \partial_x{u_z}   \nonumber\\
              \partial_z{u_y} =-\partial_y{u_z}     \nonumber\\
              \partial_z{u_z}   =-(1-2\beta^2/\alpha^2)\Big(\partial_xu_x + \partial_y{u_y} \Big) \nonumber
          \end{array}  \right. .
\label{condition_surface_libre}
\end{equation}
It follows that the divergence and curl of the wavefield can be expressed in
terms of derivatives along horizontal coordinates exclusively
\begin{align}\label{div}
\nabla \cdot \mathbf{u} = &2\frac{\beta^2}{\alpha^2}
   \left( \partial_x{u_x} + \partial_y{u_y} \right) ,  \\
 \label{curl}
\nabla \times \mathbf{u}= & \left(
2 \partial_yu_z ,  
-2 \partial_x{u_z}    ,
  \partial_x{u_y}   -\partial_y{u_x}  \right)  .
\end{align}
We now study the validity of our estimates of spatial derivatives in the data. 
Practically, we approximate $\partial_x{u_x} $ with the finite-difference formula (\ref{approx_der}). 
The estimate is based on the assumption that the two signals are very similar, i.e.,
the change of waveform between two stations is infinitesimal. 
Using the converse assumption we propose a method to determine the limit of validity of the finite
 difference approximation. When  two signals  are poorly correlated the following result
 applies: $\langle u_i(1)u_i(2)\rangle  \;\ll \;\overline{u}_i^2$. Therefore we can obtain the
following estimate of the derivative for incoherent signals
\begin{equation}\label{bad_corr}
\begin{split}
\sqrt{ \frac{\left<(u_i(2)-u_i(1))^2\right>}{d^2}}
 =&\frac{\sqrt{\langle u_i^2(2)\rangle +\langle u_i^2(1)\rangle -2\langle u_i(1)u_i(2)\rangle }}{d} \\
 \approx&\frac{\sqrt{2}\;\overline{u}_i}{d} 
\end{split}
\end{equation}
Therefore we expect to find  a $1/d$ behavior  when the signals at 2 stations are poorly correlated, i.e. sufficiently far apart.
Figure \ref{stab_der} shows the finite difference $\tfrac{u_i(\mathbf{r}_1+ d\mathbf{j})-u_i(\mathbf{r}_1)}{d}$
 as a function of inter-station distance $d$.  In the $5-7$Hz  frequency band, and 
for distances shorter than 50m, the evaluation is  unstable because the time shifts between
the two waveforms becomes nearly equal to the sampling period. For distances greater than 150m, we roughly 
find the predicted
$1/d$  behavior (see equation (\ref{bad_corr})). For inter-station distances ranging from 50m to 150m,
the finite difference curve shows a plateau, indicating that the measurement is stable and reliable.  
In  order to avoid possible biasing by extreme values that sometimes 
appear, we define the experimental derivative as the median  of all  available finite difference values.

\subsection{ Stabilization of shear to compressional energy ratios}

\begin{figure}
\begin{center}
\epsfxsize=14.5cm
\epsfbox{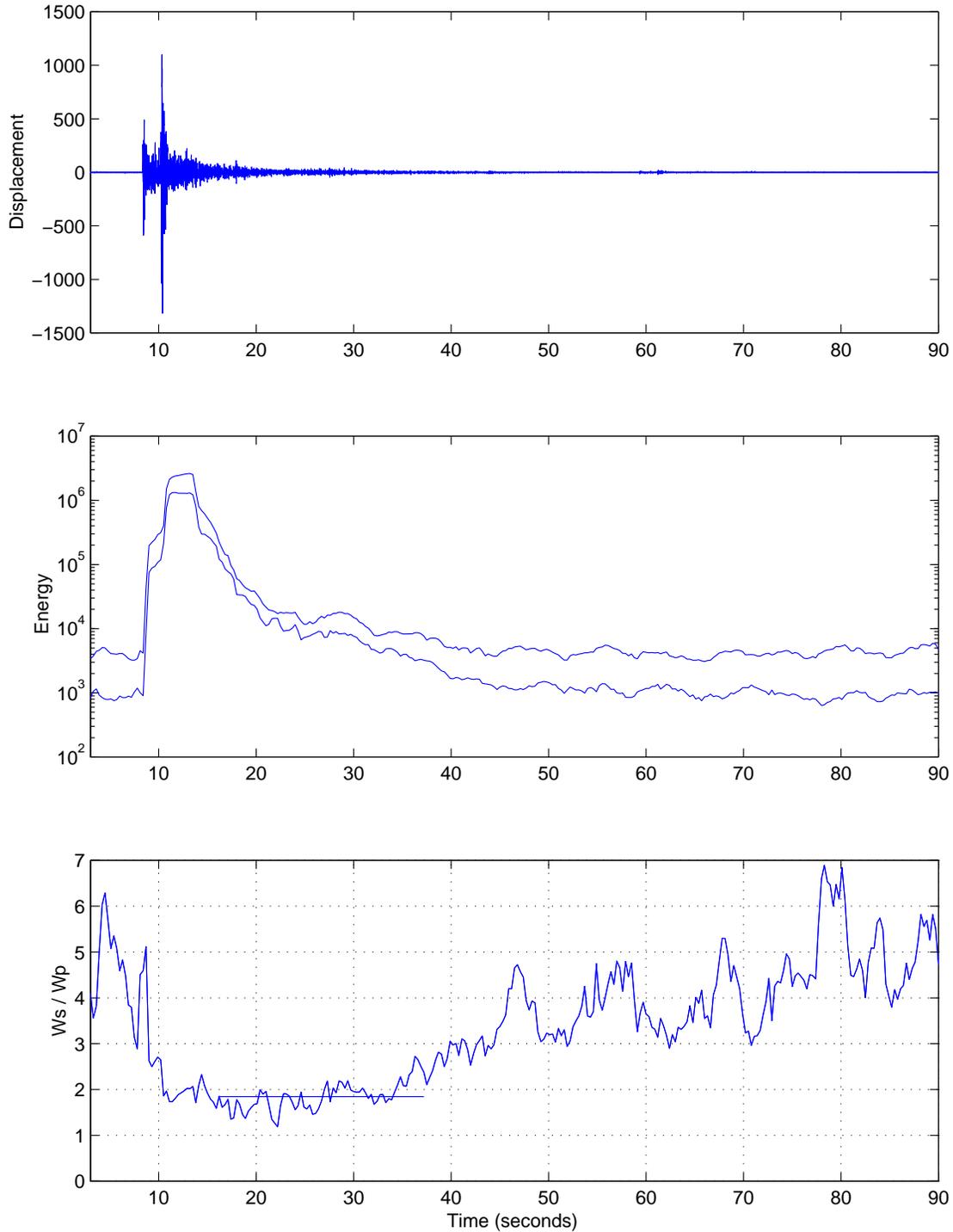}
\end{center}
\caption{ Observation of the stabilization of deformation energies  in the seismic coda.
Top~: example of vertical displacement for a shallow magnitude $2.0$ local event. Middle~: compressional 
and shear deformation energies as a function of time. 
Note the  logarithmic scale on the vertical axis. Bottom~: time dependence of the shear to compressional energy ratio. 
The ratio shows small random fluctuations while the total energy decays by two orders of magnitude. Note the
large fluctuations of energy ratios in the noise following the coda. 
The horizontal line indicates the time window used to estimate the stabilization value.}
\label{Equipartition}
\end{figure}

\begin{figure}
\begin{center}
\epsfxsize=12cm
\epsfbox{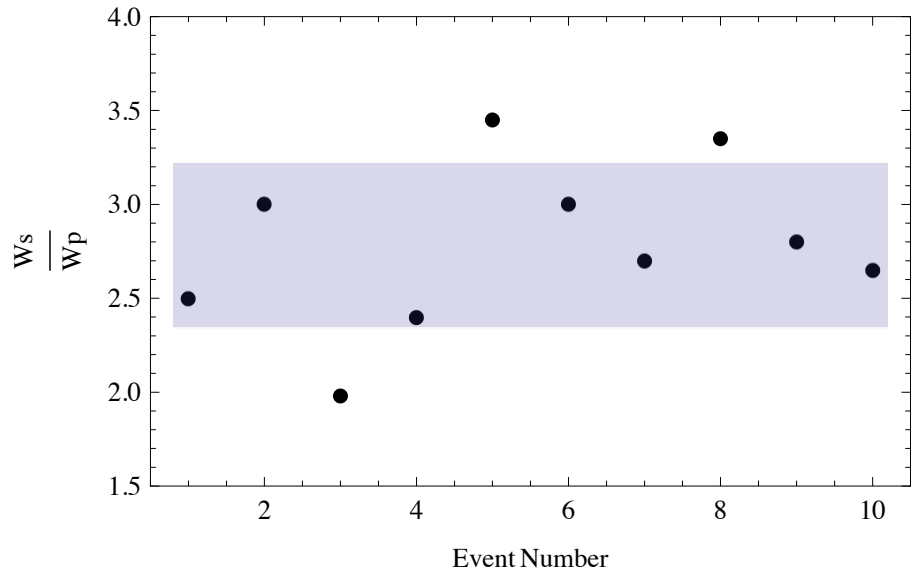}
\end{center}
\caption{Estimate of shear to compressional energy ratios for 10 local events in the $5-7$Hz frequency band. 
The stabilization ratio varies little from one event to the other. The shaded region indicates plus or minus
one standard deviation around the mean value.
This feature agrees with the theoretical prediction that the energy ratio in the coda is independent from the source parameters.}
\label{Equi_stat}
\end{figure}

Compressional ($W^p$) and shear ($W^s$) energies have been
measured for each event and for different frequency bands.  Figure \ref{Equipartition} shows a typical example of 
the band-pass filtered signal  between 5 and 7 Hz, the decay of energies with time, and the dynamics
 of the ratio between  shear and compressional energies.
For this event a stabilization of the energy ratio is clearly visible in the
coda.  While the total energy itself decreases by two orders of magnitude,
the ratio shows remarkably weak fluctuations. In addition, we infer that the
stabilization occurs very rapidly, only a few cycles after the arrival of the direct $S$
wave. When the compressional energy $W^p$  reaches the background noise level, the energy ratio again
 fluctuates randomly and stabilization  disappears. This observation indicates that the wave contents
of noise and coda are rather different. In a sense, the coda is easier to interpret because it
is likely to contain an equipartition mixture of all the modes of the system.
 Because fluctuations are weak, we define the stabilization ratio as the average of $W^s/W^p$ over 
a time window starting  one second after the direct $S$ arrival and ending when $W^p$ is of the order
of the noise level.
The equilibration has been observed for the ten earthquakes provided that frequencies lower than 3 Hz and greater than 
9 Hz have been removed. The low frequency coda is difficult to measure because of the poor sensitivity of the sensors.
 At too high frequencies the diffuse field is not sufficiently coherent to obtain  reliable estimates of the derivatives.
 The results are shown in Figure \ref{Equi_stat}, where we illustrate the fact that
the equipartition ratio is independent of the source parameters. This favors an interpretation of the 
observed stabilization as a consequence of multiple scattering. However,
the equilibration ratio equals 2.8 +/- 0.4, which is much lower than the  
ratio 7.2 predicted at the surface of a homogeneous Poisson half-space  (Figure \ref{halfspace}).
A likely explanation is the non-uniformity of velocities in the upper crust.
The seismic properties under the PFO array have been studied by \citet{fletcher90}
and \citet{vernon98}. These authors have shown that the first 50 to 70 meters
at Pinyon Flats are composed of weathered granite with low seismic velocities. 
At greater depth, intact granite with high wavespeeds is found. Therefore,
we speculate that the anomalously low energy ratio  at PFO is caused
by the presence of low-velocity layers.  We further explore this hypothesis 
by applying equipartition theory in stratified media.

\section{Modeling  energy  ratios in the coda}
\begin{figure}
\vspace*{1cm}
\begin{center}
\epsfxsize=14cm
\epsfbox{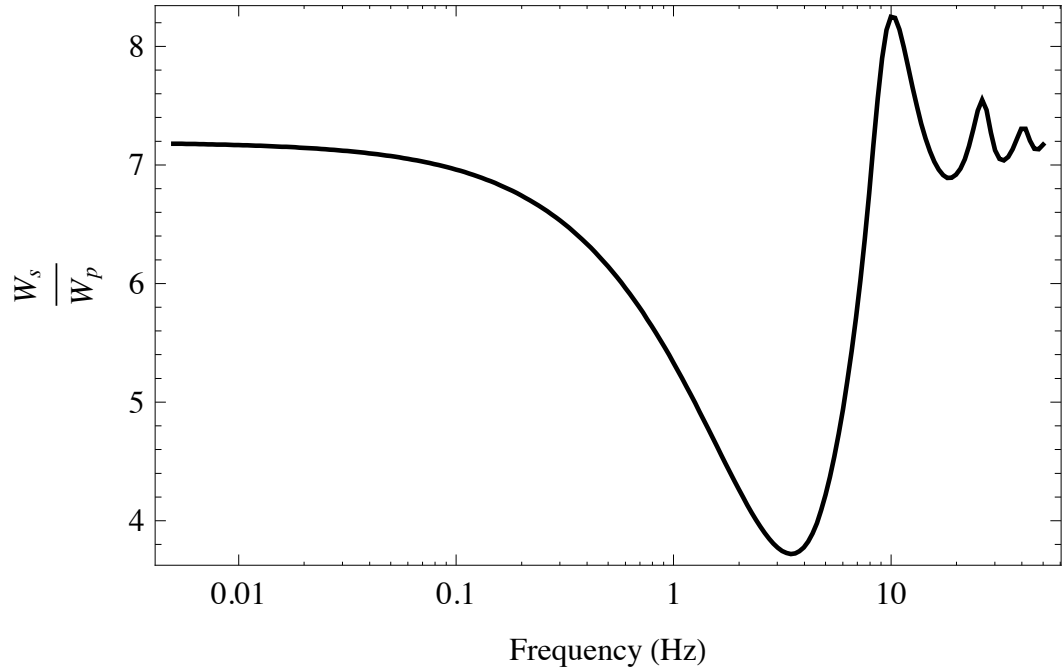}
\end{center}
\caption{ Frequency dependence of the equipartition ratio between shear ($W_s$) and compressional ($W_p$) energies  in a one-layer model
of Pinyon Flats Observatory.  At a depth of 65m, the shear velocity increases by a factor almost 2 (model 1 in Table \ref{models}). 
At low and high-frequencies,
 the equipartition ratio of a Poisson half-space is recovered. Close to the resonance frequency of the low-velocity layer,
 a significant drop of the ratio $W_s/W_p$  is observed.  }
\label{pfosimple}
\end{figure}

The structure under the PFO array was explored by  \citet{fletcher90} and \citet{vernon98}. 
We first study a simplified model  of the velocity and density profiles,
composed of a thin layer  (65 m thickness)  with relatively low velocities (2.77 km/s for $P$-wave and
1.6 km/s for $S$-wave) overlying a half-space with typical crustal velocities (5.2 km/s for $P$-wave and
3 km/s for $S$-wave). The elastic medium is poissonian everywhere and the density is uniform.
The properties of the different seismic models studied in this paper are summarized in Table \ref{models}.
In Figure \ref{pfosimple}, the  frequency dependent theoretical ratio between shear ($W^s$) and compressional ($W^p$) 
energies at equipartition is represented. At low frequency, the waves are insensitive to the 
thin low-velocity layer and the 7.2 ratio of a homogeneous Poisson half-space is recovered.
At low frequency, the fundamental Love wave can be completely neglected, and 
the  largest contribution to the  $S$ and $P$ deformation energies is made first by the Rayleigh waves,
 and second by the generalized eigenfunctions representing body waves in the lower half-space.
At very high frequency the $W^s/W^p $ ratio again converges to the half-space value. This is not
so surprising, since we may expect the  diffuse wavefield to be rather insensitive to the deep velocity contrast. 
The pronounced drop of the energy ratio to values smaller than 4 around the fundamental
resonance frequency of the layer is largely due to the change  of the Rayleigh
wave ellipticity and the increasing importance of the fundamental Love mode near the resonance frequency of the layer.
 At high frequencies,  the generalized eigenfunctions corresponding to body waves in the underlying half-space   play little role in the equipartition
ratio. The high-frequency oscillations are mostly due to the interplay between Rayleigh and Love  waves.

\begin{figure}
\vspace*{1cm}
\begin{center}
\includegraphics[width=0.7\linewidth]{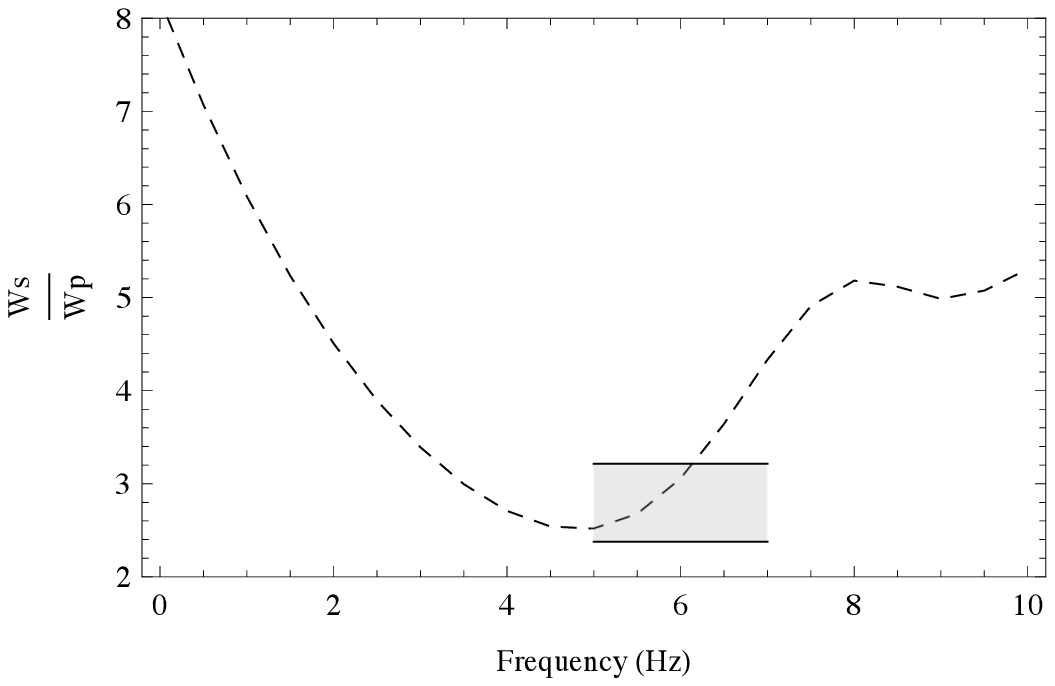}
\end{center}

\vspace*{0.4cm}
\begin{center}
\includegraphics[width=0.7\linewidth]{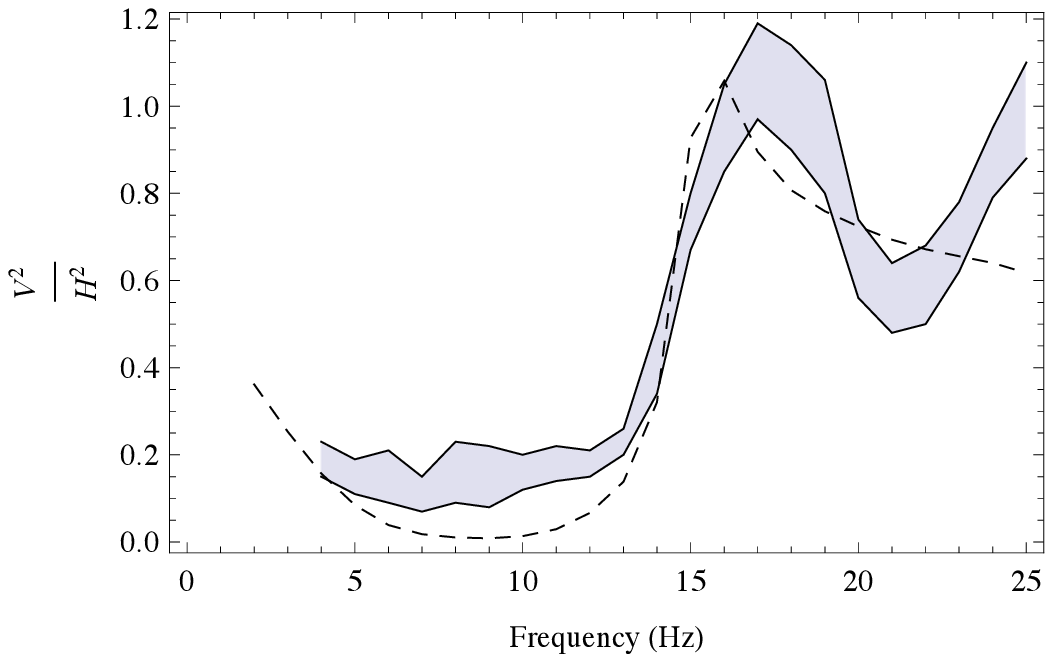}
\end{center}
\caption{Modeling of the measured stabilization ratios with a simple 2-layer model of PFO as described in
Table \ref{models} (model 2). Top: Observed (shaded region) and modeled (dashed line) shear to compressional energy ratio. 
 Bottom: Observed (shaded region) and modeled (dashed line) vertical to horizontal kinetic energy ratio.  The black solid lines
delimit the $\pm 1$ standard deviation region around the mean value.
Agreement is reasonable up to
 15 Hz. }
\label{best2layer}
\end{figure}

Because the deformation energies are accessible in a narrow frequency band only,
we have also evaluated the ratio of vertical ($V^2$) to horizontal ($H^2$) kinetic
energies in the 5-25 Hz band. The measurement of this ratio is straightforward and offers additional
information on the local structure. The experimental results
are  plotted in Figure \ref{best2layer} and show
a clear frequency dependence. The ratio decreases from 0.5 at frequencies around 4 Hz
to about 0.1-0.2 in the 6-12Hz frequency band, then rapidly increases to 0.8,
 and oscillates around this value in the 15-25Hz frequency band.   
While a simplified 1-layer  model (model 1 in Table \ref{models}) gives a clear qualitative explanation 
 of the observations at PFO,  it is not quantitatively  satisfactory.
By a process of trial and error and using  a-priori information provided by 
previous studies, we tried to find 2-layer models that better
match the observation.  The velocity and density profile of a typical  2-layer model
is given in Table \ref{models}. The velocities in the bedrock and
in the deepest part of the weathered zone are identical to those given by \citet{fletcher90}. 
According to these authors the near-surface seismic properties are more variable and less well
 constrained. Therefore we explored various thicknesses and velocities for 
the top layer. Regarding the  density, little information could be found in the literature.
We adopted a typical value of 2.7 for the  intact bedrock and allowed the density to
decrease to 2.2  in the weathered, less consolidated layers. 
Observed and theoretical energy ratios for a 2-layer model that fits
 reasonably  well the $W^s/W^p$ and $V^2/H^2$ ratios up to 15 Hz are shown in Figure \ref{best2layer}.
In the simple 2-layer model, the small $V^2/H^2$ ratio at low frequency is due simultaneously
to the nearly horizontal polarization of the  fundamental mode Rayleigh wave  and the increasing role played by the
fundamental Love wave. For a thorough study of the frequency-dependent ellipticity of the Rayleigh wave in layered structures,
the reader is referred to \citet{malischewsky04}.
 In model 2 (see Table \ref{models}), at frequencies higher than 5 Hz, the surface waves
trapped in the low-velocity layers largely dominate over body waves coming from below the structure. 
The sharp increase of the $V^2/H^2$ ratio at high-frequencies is again explained by an interplay
between Love and Rayleigh waves. Around 15 $Hz$, the fundamental mode
Rayleigh wave becomes strongly vertically polarized and the weight of Love waves in the equipartition
ratio suddenly drops. The two effects combined explain the peak value observed in the data.
Calculations based on Rayleigh waves only, predict fluctuations of the $V^2/H^2$ ratio
much bigger than observed. This demonstrates that the contribution of Love waves is of fundamental importance.
At frequencies higher than 15Hz, the measured kinetic energy ratio oscillates and departs significantly
from the model calculations. This may  indicate a departure from the simple layered structure.

\begin{figure}
\vspace*{1cm}
\begin{center}
\includegraphics[width=0.7\linewidth]{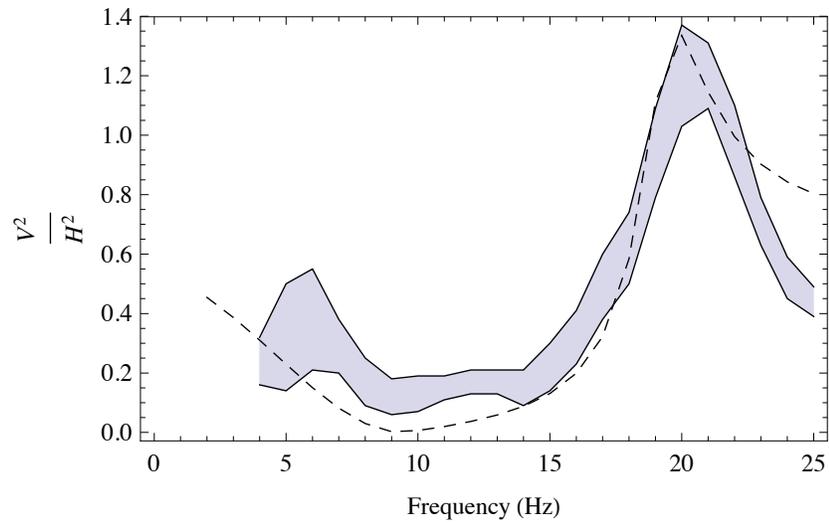}
\end{center}
\caption{Modeling of the measured stabilization ratio between vertical and horizontal kinetic energies
  with a  3-layer model of PFO as described in
Table \ref{models} (model 3). The shaded area delimits the  $\pm 1$ standard deviation region for the data. The dashed line shows
the model calculations.  
}
\label{siteeffect}
\end{figure}

 By exploring
a large number of models, we found that the position of the steep rise from the low to high frequency behavior is extremely
sensitive to the the thickness and velocity of  the upper layer. This is illustrated in Figure \ref{siteeffect} where the
frequency dependent vertical to horizontal kinetic ratio has been measured at one station located at the end
of one of the two orthogonal arms (see Figure \ref{fig:carte}). Figure \ref{siteeffect} also shows the calculations
of the frequency dependent ratio between the vertical and horizontal kinetic energies for model 3, Table \ref{models}.
 This model is completely similar to model 2,  except in the first 15 meters where the velocities are
slightly different from those proposed by \citet{fletcher90}. 
This model fits reasonably well the observations over the whole frequency band. 
The comparison of Figures \ref{best2layer}
 and \ref{siteeffect} demonstrates that the frequency dependence of the ratio
between vertical and horizontal kinetic energies measured in the coda  contains  information on the local velocity structure. 
In the case of PFO, our study supports the idea that the velocity profile  presents
 simultaneously velocity gradients at depth and thin very low-velocity layers at the surface.

\section{Conclusion and outlook}
We have observed the stabilization of ratios of deformation and kinetic energies
in the coda of small earthquakes, at PFO. The stabilization phenomenon
is interpreted as a sign of multiple scattering of waves in the crust at high frequencies.
The ratio between the   shear and compressional energies  can be accurately estimated in the 
5-7 Hz frequency band and was observed to be much lower than the one found in Mexico by \citet{shapiro00}. 
To understand this observation, we have developed a theory of equipartition in arbitrary 
layered elastic media based on the spectral decomposition of the elastodynamic operator.
We have shown that the decrease of the $S$ to $P$ deformation energy ratio may be explained by the low-velocity
sub-surface  layers. 
Like in Mexico \citep{shapiro00,hennino01}, the 
stabilization of various energy ratios occurs shortly  after
the {\it S} wave arrival. Such a rapid stabilization has also been found
in numerical simulations \citep{margerin00}, typically after a moderate number
(3-4) of scattering events.   This suggests that the mean free time
 -the typical time between two scattering events- is very small, of the order
of a few seconds only, and supports the idea of a highly heterogeneous crust in California.
Our results on the frequency dependence of the ratio of  vertical to horizontal   kinetic energies
sound reminiscent of the so-called Nakamura's technique which relies on 
the frequency dependence  of the $H/V$ spectral ratio of ambient noise vibrations
 \citep[see ][for a review]{bard98}. This technique usually
makes use of the low-frequency  $H/V$ peak to retrieve the resonance frequency
of sedimentary deposits. The physical bases and limitations of the  $H/V$ method are still 
actively debated. \cite{bonnefoy06} present a detailed study of a simple 1-layer configuration
which illustrates the complex interpretation of $H/V$ measurements depending on the spatial distribution of the noise sources. 
Because coda waves are composed of an equipartitioned mixture of all
modes, they are in a sense easier to model than noise signals.
 We have shown that it is possible  to model reasonably well the frequency dependent
ratio of  vertical to horizontal kinetic energies  in the coda at PFO. In addition to a low-frequency
 global resonance of the 1-D structure, we find that some information on the uppermost layers can also be obtained. 
Another interesting aspect of the coda is the possibility to check the wave content of the signal through
the measurements of several energy ratios that can be compared to theoretical predictions. 
Further investigations of site effect assessments with coda waves are required before reaching
more definitive conclusions. 

\clearpage
\bibliography{bibpfo}
\clearpage

\begin{table}
  \begin{center}
    \begin{tabular}{|c|c|c|c|c|c|}
      \hline
      Origin time & Latitude & Longitude & Magnitude & Depth & Source-array
      distance(km) \\
      \hline
      117.12.52.42.392 & 34.0487 & -116.3908 & 2.3 & -2.2 & 48.89 \\
      \hline
      118.08.32.38.165 & 33.8647 & -116.1659 & 2.4 & 5.1 & 39.04 \\
      \hline
      122.11.35.03.691 & 33.4927 & -116.4624 & 2.1 & 7.0 & 13.20 \\
      \hline
      125.08.10.15.667 & 33.5074 & -116.4672 & 2.0 & 8.3 & 11.64 \\
      \hline
      127.12.40.57.581 & 33.8693 & -116.1555 & 2.0 & 3.7 & 40.08 \\
      \hline
      130.07.23.35.987 & 33.6496 & -116.7291 & 2.1 & 15.7 & 25.46 \\
      \hline
      130.14.25.10.270 & 33.1944 & -116.3622 & 2.8 & 14.5 & 47.13 \\
      \hline
      132.23.54.49.190 & 33.9839 & -116.3103 & 2.5 & 4.8 & 43.50 \\
      \hline
      134.05.05.27.123 & 33.5384 & -116.6073 & 2.6 & 11.8 & 16.01 \\
      \hline
      136.01.14.21.003 & 33.4376 & -116.4491 & 2.4 & 11.1 & 19.30 \\
      \hline
    \end{tabular}
  \end{center}
 \caption{Location and magnitude of the 10 events used in this study.}
\label{sourceparam}
\end{table}

\clearpage
\begin{table}
\begin{center}
\begin{tabular}{|c|c|c|}
\hline
 Model 1 & Model 2 & Model 3 \\
\hline
                           &                             &$ h_1 =4$m   \\
                           &                             & $\alpha_1 =300$m/s \\
                           &                             & $\beta_1 =150$m/s \\
                           &    $h_1 = 11$ m              &  $\rho_1 =2200$kg/m$^3$ \\
                           &    $\alpha_1 =720$m/s         &                     \\  
  $h_1 =65 $m              &    $\beta_1 =  400$m/s         &    $ h_2 =11 $m   \\
  $\alpha_1 =2770$m/s      &    $\rho_1 = 2200$kg/m$^3$  & $\alpha_2 =900 $m/s \\
  $\beta_1 =1600$m/s       &                             & $\beta_2 = 500$m/s \\
  $\rho_1 = 2700$kg/m$^3$  &    $h_2 = 41$ m             &  $\rho_2 = 2200$kg/m$^3$ \\
                           &    $\alpha_2 =3100$m/s          &                          \\ 
                           &    $\beta_2 = 1600$m/s          & $ h_3 = 50$m   \\
                           &    $\rho_2 = 2500$kg/m$^3$     & $ \alpha_3 =3100 $m/s \\
                           &                                &  $\beta_3 =1600 $ m/s \\
                           &                                &  $\rho_3 = 2700$kg/m$^3$ \\
\hline
   $\alpha_{\infty} = 5200$m/s    &   $\alpha_{\infty}=5400$m/s         &  $\alpha_{\infty} = 5400$m/s         \\ 
   $\beta_{\infty} = 3000$m/s     &   $\beta_{\infty}=3000$m/s          & $\beta_{\infty}  = 3000$m/s     \\
 $\rho_{\infty} =2700$ kg/m$^3$    &   $\rho_{\infty}=2700$kg/m$^3$     & $\rho_{\infty} = 2700 $kg/m$^3$   
\end{tabular}
\vspace*{0.5cm}
\caption{Seismological models of the sub-surface at Pinyon Flats Observatory used in this study.
 The different depths  and velocities are inspired by the 
 results of \protect\citet{fletcher90}.}
\label{models}
\end{center}
\end{table}

\clearpage

\clearpage

\clearpage

\clearpage

\clearpage

\clearpage

\clearpage

\clearpage


\clearpage

\end{document}